\documentclass[a4paper,fleqn,usenatbib]{mnras}

\usepackage{newtxtext,newtxmath}

\usepackage[T1]{fontenc}
\usepackage{ae,aecompl}

\usepackage{graphicx}	
\usepackage{amsmath}	
\usepackage{amssymb}	

\title[Evolution of Sr and Ba]{The contribution from rotating massive stars to the enrichment in Sr and Ba of the Milky Way}

\author[Rizzuti et al.]{ Rizzuti, F.$^{1}$, Cescutti, G.$^{2}$, Matteucci, F. $^{1,2,3}$,
  Chieffi, A.$^{4,5}$, Hirschi, R.$^{6,7}$, Limongi, M.$^{7,8}$
\thanks{E-mail: email}
\\
$^{1}$Dipartimento di Fisica, Sezione di Astronomia, Universit\`a degli Studi di Trieste, Via Tiepolo 11, I-34143 Trieste, Italy\\
$^{2}$INAF, Osservatorio Astronomico di Trieste, Via Tiepolo 11, I-34143 Trieste, Italy\\
$^{3}$INFN, Trieste, Via Valerio 2, I-34127 Trieste, Italy\\
$^{4}$INAF/IAPS, Via Fosso del Cavaliere 100, I-00133 Roma, Italy\\
$^{5}$Monash Centre for Astrophysics (MoCA), School of Mathematical Sciences, Monash University, Victoria 3800, Australia\\
$^{6}$Astrophysics Group, Lennard-Jones Laboratories, Keele University, Keele ST5 5BG, UK\\
$^{7}$Kavli IPMU (WPI), The University of Tokyo, Kashiwa, Chiba 277-8583, Japan\\
$^{8}$INAF/Osservatorio Astronomico di Roma, Via di Frascati 33, I-00040 Monte Porzio Catone, Italy}

\date{Accepted XXX. Received YYY; in original form ZZZ}

\pubyear{2018}

\begin{document}
\label{firstpage}
\pagerange{\pageref{firstpage}--\pageref{lastpage}}
\maketitle

\begin{abstract}
Most neutron capture elements have a double production by r- and s-processes, but the question of production sites is complex and still open. Recent studies show that including stellar rotation can have a deep impact on nucleosynthesis.
We studied the evolution of Sr and Ba in the Milky Way. A chemical evolution model was employed to reproduce the Galactic enrichment. We tested two different nucleosynthesis prescriptions for s-process in massive stars, adopted from the Geneva group and the Rome group. Rotation was taken into account, studying the effects of stars without rotation or rotating with different velocities. We also tested different production sites for the r-process: magneto rotational driven supernovae and neutron star mergers.
The evolution of the abundances of Sr and Ba is well reproduced. The comparison with the most recent observations shows that stellar rotation is a good assumption, but excessive velocities result in overproduction of these elements. In particular, the predicted evolution of the [Sr/Ba] ratio at low metallicity does not explain the data at best if rotation is not included. Adopting different rotational velocities for different stellar mass and metallicity better explains the observed trends.
Despite the differences between the two sets of adopted stellar models, both show a better agreement with the data assuming an increase of rotational velocity toward low metallicity. Assuming different r-process sources does not alter this conclusion.

\end{abstract}




\section{Introduction}
It has been known for a long time, since \citet{Burbidge}, that
elements heavier than $^{56}$Ni, which has the maximum binding energy
per nucleon, are mostly synthesized by neutron capture. Depending on
the timescale of neutron capture compared to the $\beta$ decay: the
slow process (s-process) when the neutron capture takes place on
longer timescale, the opposite for rapid process (r-process).
\\Concerning the s-process, major production sites have been identified in
low-mass asymptotic giant branch (AGB) stars (mass range 1.5-3.0
$M_\odot$) \citep{Cristallo09,Cristallo11,Karakas10}. AGB stars can
produce all the neutron capture elements up to Pb and Bi and the main
source of neutrons in this case is the reaction $^{13}$C($\alpha$,n)$^{16}$O. Massive stars can also produce neutron capture elements with an
s-process. In massive stars, the neutron flux is weaker and is
originated in this case by the reaction $^{22}$Ne($\alpha$,n)$^{25}$Mg.  The
s-process production is called in this case ``weak s-process'' and
typically the lower neutron flux does not allow this process to built
up very heavy elements, but only elements up to the magic number 50 such as Sr-Y-Zr. The first calculations of the 90's
\citep{Raiteri92} showed a strong metal dependency, and basically at
metallicity lower than a tenth of the solar no production was
expected. This has also been confirmed by \citet{Limongi03}
on the basis of a larger grid of initial masses and metallicities. For this reason, the s-process production by
massive stars at extremely low metallicity did not have an impact in
previous chemical evolution models
\citep{Trava99,Trava04,Cescutti06}. However, recent stellar
evolution studies showed that rotation-induced mixing may keep in contact regions otherwise separate in absence of rotation. Such a phenomenon induces a peculiar nucleosynthesis as well as an increase of the nuclear burning cores, of the stellar lifetimes and of the amount of mass lost during the evolution (see \citealt{Chieffi13}). Interesting
implications are found at low metallicities, where stars are expected
to be more compact and rotate faster, intensifying the effects
(\citealt{Meynet}; \citealt{Fris16}, and more recently
\citealt{Lim18}). In terms of chemistry, one of the results of rotating mixing for massive stars is an enhancement of nitrogen and s-process
production of neutron capture elements.  Any investigation on the
impact of this s-process production by massive stars has to deal with
the production of neutron capture elements by r-process events. The r-process requires an extremely neutron-rich
environment.  Before the event GW170817 \citep{Abbott17}, it was
unclear where in nature the r-process can take place and several sites
were proposed, and also now we cannot conclude that neutron star mergers are the only
r-process events in nature \citep{Cote18,Simonetti19}.  Core-collapse
SNe or electron capture SNe were certainly the first proposed sites
\citep{Truran, Cowan}. Following theoretical studies \citep{Arcones07}
found that these sites do not have proper entropy and neutron fraction
to have an efficient r-process activation.  Therefore alternative
sites and mechanisms were proposed, in replacement of or in addition to
SNe: neutron star mergers (NSMs) \citep{Ross} or magneto-rotationally
driven supernovae (MRD SNe) \citep{Winteler,Nishi}.  \\A study of the
chemical evolution enrichment adopting NSMs as source of r-process
material have been carried on by \citet{Matteucci} (so before the NSM
event GW170817 observed by LIGO and Virgo \citep{Abbott17}).  The
conclusions were that NSMs may be responsible for the r-process
enrichment in the Galactic halo either totally or just in part, in a
mixed scenario with both SNe II and NSMs, providing a very short
time-scale for the merging after the formation of the neutron star
binary \citep[see also][]{Arg04,Cescutti15,Simonetti19}. Similar
studies by means of chemical evolution models have been carried on for
MRD SNe by \citet{Cescutti14}, whereas in \citet{Cescutti13} the EC
SNe scenario was investigated.
\citet{Cescutti13}, \citet{Cescutti14}, \citet{Cescutti15} were the first studies
showing that s-process driven by rotation in massive stars has a
fundamental role for chemical evolution results.  Independently by the
r-process event considered, the s-process production by massive stars
was shown to be a possible solution to explain a signature in neutron capture
elements in the Galactic halo: the spread in light (e.g. Sr-Y-Zr) to
heavy neutron capture elements (e.g. La, Ba). These results were
obtained adopting a nucleosynthesis based on
\citet{Fris12,Fris16}. More recently, \citet{Prantzos} obtained
similar results, but using the theoretical nucleosynthesis obtained by
\citet{Lim18}. \\The main purpose of this paper consists in testing
and comparing nucleosynthesis prescriptions for rotating massive stars
using these two studies, \citet{Fris16} and \citet{Lim18}.  The aim
is to analyse the effects produced by prescriptions coming from
different assumptions on stellar evolution.  Concerning the r-process
component, we assume two scenarios, the MRD SNe scenario using the
prescriptions obtained thanks to chemical evolution models by
\citet{Cescutti14} and NSMs using the prescriptions from
\citet{Matteucci}.  The tool we use for this study is a chemical
evolution model of the Galaxy, based on the two-infall model \citet{Chiappini}. The resulting
abundances of elements - we will focus on Sr and Ba - as functions of
metallicity are compared to the observations, to verify if the
prescriptions assumed reproduce the data correctly, and if different
works show compatible results.  The paper is organized as follows: in Section 2 we describe
the adopted observational data. In Section 3 the chemical evolution
model is presented. In Section 4 the adopted nucleosynthesis
prescriptions are discussed. In Section 5 the results are presented
and in Section 6 some conclusions are drawn.

\section{Observational data}
Three main sources of chemical abundances of Galactic stars were
adopted. For low metallicities ([Fe/H] from $-4$ to $-1$) Milky Way
halo stars abundances were taken from various authors (JINA-CEE
database). The totality of the authors is displayed in
Table~\ref{tab:1}.  \\Another sample, from \citet{Batt}, was used for
the data belonging to the thin and thick disks of the Milky Way. In
their paper, the original data from \citet{Bensby} were extended to
include also the abundances of neutron capture elements.  
\\Finally, the data of the halo star TYC 8442-1036-1 from the work of \citet{Cescutti16} were taken into account ([Fe/H] $=-3.5$).
\\All the
authors normalized the data according to solar abundances taken from
\citet{Asp}.
\begin{table*}
\centering
\footnotesize
\caption{Sources for observational data abundances.}\label{tab:1}
\begin{minipage}{0.3\linewidth}
\begin{tabular}{lcc}
\hline
\hline
 & Ba & Sr \\
\hline
\citet{Aoki02}
& X & X \\
\citet{Aoki05}
 & X & X \\
\citet{Aoki07b}
& X & X \\
\citet{Aoki07a}
& X &  \\
\citet{Aoki08}
& X & X \\
\citet{Aoki12}
 & X &  \\
\citet{Aoki13}
& X & X \\
\citet{Aoki14}
& X & X \\
\citet{Bark05}
& X & X \\
\citet{Batt}
& X & X \\
\citet{Bensby11} 
& X &    \\
\citet{Bonif09} 
& X & X \\
\citet{Burris00} 
& X & X \\
\citet{Caffau11} 
& X & X \\
\citet{Carretta02} 
& X & X \\
\citet{Cayrel04}
& X & X \\
\citet{Cescutti16}
& X & X \\
\citet{Christlieb04}
& X & X \\
\citet{Cohen04}
& X & X \\
\citet{Cohen13}
& X & X \\
\citet{Cowan02}
& X & X \\
\citet{Frebel07}
& X &    \\
\citet{Fulbright00}
& X &        \\
\citet{Hansen12}
& X & X \\
\citet{Hansen15}
& X & X \\
\citet{Hayek09}
& X & X \\
\citet{Hollek11}
& X & X \\
\citet{Honda04}
& X & X \\
\citet{Honda11}
& X & X \\
\citet{Ishigaki10}
& X &         \\
\citet{Ishigaki13}
& X & X \\
\hline
        \end{tabular}
    \end{minipage}%
    \begin{minipage}{0.3\linewidth}
    \begin{tabular}{lcc}
\hline
\hline
 & Ba & Sr\\
\hline
\citet{Ivans03}
& X & X \\
\citet{Ivans06}
& X & X \\
\citet{Jacobson15}
& X & X \\
\citet{Johnson02}
& X & X \\
\citet{Jonsell05}
& X &                  \\
\citet{Lai07}
& X & X \\
\citet{Lai08}
& X & X \\
\citet{Li15a}
& X & X \\
\citet{Li15b}
& X & X \\
\citet{Mashonkina10}
& X & X \\
\citet{Mashonkina14}
& X & X \\
\citet{Masseron06}
& X & X \\
\citet{Masseron12}
& X &        \\
\citet{McWilliam95}
& X & X \\
\citet{Norris97a}
& X & X \\
\citet{Norris97b}
& X & X \\
\citet{Norris97c}
& X & X \\
\citet{Placco14}
& X & X \\
\citet{Placco15}
& X & X \\
\citet{Preston00}
& X & X \\
\citet{Preston06}
& X & X \\
\citet{Roederer10}
& X & X \\
\citet{Roederer14}
& X & X \\
\citet{Ryan91}
& X & X \\
\citet{Ryan96}
& X & X \\
\citet{Siq14}
& X & X \\
\citet{Spite14}
& X & X \\
\citet{Westin00}
& X & X \\
\citet{Yong13}
& X & X \\
\citet{Zhang09}
& X &     \\
\\
\hline
\end{tabular}
    \end{minipage} 
\end{table*}

\section{The chemical evolution model}
The model employed reproduces the evolution of the Galaxy, assuming
one main infall episode. It is based on the two-infall model
\citet{Chiappini}, but it has only one single infall. The choice of a
sequential one-infall model is based on the consideration that the
main difference with the two-infall model is the gap in star formation
at the end of the halo-thick disk phase, which is not yet proven
observationally. On the other hand, the main behavior of the abundance
ratios as functions of [Fe/H] is very similar. Moreover, here we do
not distinguish between thick and thin disk stars. The adopted model
is a homogeneous one, namely it assumes instantaneous mixing
approximation: the gas mixing time is considered smaller than the
timestep of integration, so the ISM is well mixed at all times. The
values are calculated for one zone, the solar vicinity. We assume the
age of the Milky Way to be 14 Gyr.  \\The equations which rule the gas
fraction $G_i$ of a certain element $i$ are the following:
\begin{equation}
\begin{split}
 \dot{G_i}&(r_\odot,t)=-\psi(r_\odot,t)\ X_i(r_\odot,t)\ 
\\&+\int\limits_{_{M_L}}^{_{M_{Bm}}} \psi(r_\odot,t-\tau_m)\ Q_{mi}(t-\tau_m)\ \phi(m)\ \textrm{d}m
\\&+A \int\limits_{_{M_{Bm}}}^{_{M_{BM}}} \phi (m) \cdot \left[\ \int\limits_{_{\mu_m}}^{_{0.5}} f(\mu)\ \psi(r_\odot,t-\tau_{m2})\ Q_{mi}(t-\tau_{m2})\ \textrm{d}\mu \right] \textrm{d}m
\\&+(1-A)\int\limits_{_{M_{Bm}}}^{_{M_{BM}}} \psi(r_\odot,t-\tau_m)\ Q_{mi}(t-\tau_m)\ \phi(m)\ \textrm{d}m
\\&+\int\limits_{_{M_{BM}}}^{_{M_{U}}} \psi(r_\odot,t-\tau_m)\ Q_{mi}(t-\tau_m)\ \phi(m)\ \textrm{d}m
\\&+\dot{G_i}(r_\odot,t)_{inf}\ 
\end{split}
\end{equation}
where the first term on the right-hand side represents  the gas subtracted by the ISM and locked into stars, the integrals refer to the production and restitution of the element $i$ from the stars into the ISM, and the last term is the gas accretion rate. In particular, $X_i$ is the abundance by mass of the element $i$, $Q_{mi}$ the fraction of mass restored by a star of the mass $m$ in the form of the element $i$, $f(\mu)$ the distribution of the mass ratio for the secondary in a binary system giving use to SNe Ia, $\phi(m)$ the initial mass function (IMF), which here is the one suggested by \citet{Scalo}.
\\The star formation rate (SFR) $\psi(r,t)$ is expressed as:
\begin{equation} \tag{2}
\psi(r,t)= \nu \left(\frac{\Sigma(r,t_G)}{\Sigma(r,t)}\right)^{k-1} G^k_{gas}(r,t)
\end{equation}
where $\nu$ is the star formation efficiency, here 1.2 Gyr$^{-1}$, $\Sigma(r,t)$ the total surface mass density, $t_G=14$ Gyr the age assumed for the Milky Way, $k=1.4$ the law index, and $G_{gas}(r,t)$
the surface density normalized to the present time total surface mass density in the disk. The equation is the same as in \citet{Chiappini}, but the ratio $\left({\Sigma(r,t)}/{\Sigma(r_\odot,t)}\right)^{2k+1}$ in the original formula, namely the total surface mass density over the total surface mass density at the solar position, is set equal to 1 for we are in the solar neighborhood.
\\The four integrals have different meanings: 
 \begin{enumerate} 
 \item the first represents the stars in the mass range $M_L$ (0.8 $M_\odot$, the lower mass limit) to $M_{Bm}$ (3 $M_\odot$) with a lifetime of $\tau_m$; 
  \item the second represents the SNe Ia originating from white dwarfs - binary systems, from the minimum $M_{Bm}$ (3 $M_\odot$) to the maximum $M_{BM}$ (16 $M_\odot$) allowed for the whole binary systems, the parameter $A=0.05$ being the fraction of binary systems with the right characteristics to give rise to SNe Ia \citep{Matteucci01}, and $\tau_{m2}$ the lifetime of the secondary star; 
 \item the third describes the single stars with masses in the range $M_{Bm}$ to $M_{BM}$, which can end their lives either as white dwarfs (3-9 $M_\odot$) or as core collapse SNe (9-16 $M_\odot$);
 \item the fourth represents the core collapse SNe, from the minimum mass $M_{BM}$ to the maximum one $M_U$ (100 $M_\odot$, the upper mass limit).
\end{enumerate}
Finally, for the gas infall the following law is adopted:
\begin{equation} \tag{3}
\dot{G_i}(r,t)_{inf} = a(r)\ (X_i)_{inf}\ e^{-t/\tau(r)}
\end{equation}
where the parameter $a(r)$ is the total surface mass density at the present time, fixed to 65 $M_\odot\ pc^{-2}$, $(X_i)_{inf}$ are the abundances for the infalling material, assumed to be primordial, and $\tau(r)$ the characteristic time of formation for the disk, which in the solar neighbourhood is $\tau(8\ kpc)=7$ Gyr  \citep{Chiappini}.

\section{Nucleosynthesis prescriptions}
\label{sec:sec4}

\begin{figure}
\centering
\footnotesize
 \includegraphics[width=\columnwidth]{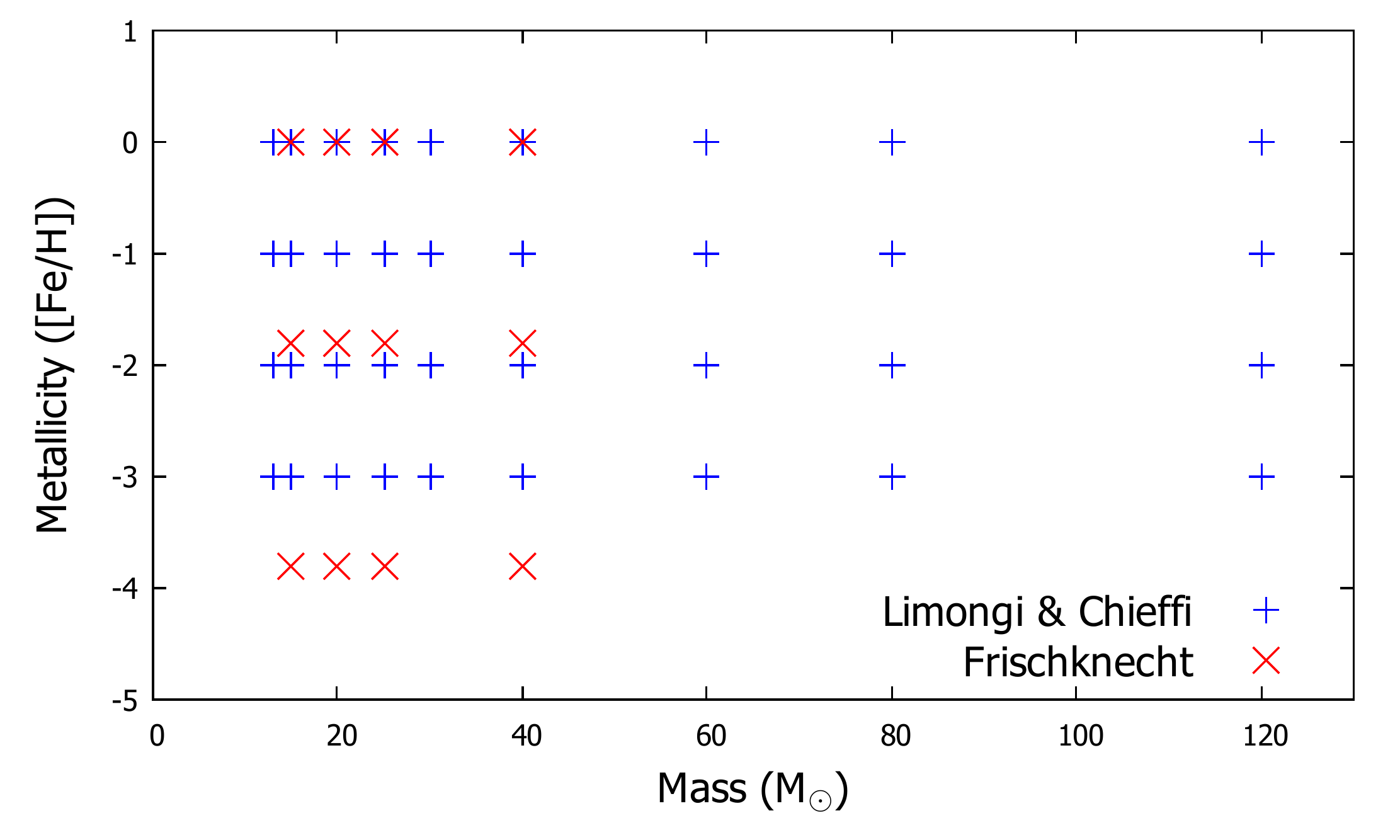}
 \caption{The two papers used for rotating massive stars \citet{Lim18} (blue crosses) and \citet{Fris16} (red crosses) compared for mass and metallicity employed in the models.}
 \label{fig:comp}
\end{figure}

As reported in the Introduction, most of the neutron capture elements
have a double production and are formed both by r-process and
s-process. This is the case of barium and strontium, the two elements
analysed in this work. The duality of production for barium has been
studied in chemical evolution models since
\citet{Trava99} and \citet{Cescutti06}, showing that Ba has a main s-process
component produced by low mass AGB stars but also an r-process
production. 
Here we assume also the s-process contribution from massive stars, taking into account the rotational velocity and stellar metallicity as intial parameters. This has already been considered in \citet{Cescutti13}, \citet{Cescutti14}, \citet{Cescutti15}, \citet{Prantzos} but here we compare for the first time two nucleosynthesis sets \citep{Fris16,Lim18} for rotating massive stars.
\\The set of nucleosynthesis for different sources of neutron capture elements is rather complex and was taken from different authors; we have summarized them in Table~\ref{tab:2}.
\begin{table*}
\centering
\footnotesize
\caption{Authors used for nucleosynthesis prescriptions.}\label{tab:2}
\begin{tabular}{lcc}
\hline
\hline
&s-process&r-process\\
\hline
Low mass stars & \citet{Cristallo09,Cristallo11} & --- \\
\\
Massive stars & \citet{Fris16} & \citet{Ross} \\
& \citet{Lim18} & \citet{Winteler} \\
\\
NSM & --- & \citet{Matteucci} \\
& & \citet{Cescutti15} \\
\hline
\end{tabular}
\end{table*}
\\Low mass stars, in a mass range of 1.3-3 $M_\odot$, which are
responsible for part of the s-process, were taken from
\citet{Cristallo09,Cristallo11}. We used results from models of
non-rotating stars. However, these non-rotating yields tend to
overproduce the neutron capture elements at solar abundance. On the
other hand, rotating yields produce significantly lower amount of
neutron capture elements. For this reason, we decided to divide by a
factor of two the non-rotating yields, since
rotational yields would have produced a similar decrease.
\\For the s-process component in
rotating massive stars, we have decided to investigate two
available dataset, \citet{Fris16} and \citet{Lim18}.
\\\citet{Fris16} studied the impact of rotation in massive stars
nucleosynthesis; they produced a large grid of yields using several
models with different features. They took into account a stellar mass
range of 15-40 $M_\odot$; the stars are assumed to have different
initial rotational velocities, selected by mass and metallicity.  Four
metallicities, expressed in [Fe/H], are explored: 0, $-1.8$, $-3.8$
and $-5.8$. In our model we considered only the first three
metallicities. For the lowest metallicitity (i.e. [Fe/H] $=-5.8$),
only a model of 25 $M_\odot$ has been computed, and for this reason we decide not to extrapolate the results. Instead, we assumed the yields from [Fe/H] $=-3.8$ also for lower metallicities.
\\For rotational scenarios in the first two metallicities [Fe/H] $=0$ and $-1.8$, we used the models for which \citet{Fris16} fixed the value of standard initial
rotation rate over critical velocity to $v_{\text{ini}}/v_{\text{crit}}=0.4$. With this parameter assumed as a constant, the
resulting average equatorial rotation velocity on the MS, $\langle v \rangle_{\text{MS}}$, increases with decreasing metallicity; for
15-20 $M_\odot$ stars at solar metallicity, $\langle v \rangle_{\text{MS}}$
corresponds to 200-220 km/s.
\\For the metallicity [Fe/H] = $-3.8$, in order to include a stronger production of s-process, we decided to use the model providing a faster rotation ($v_{\text{ini}}/v_{\text{crit}}=0.5$) and a lower $^{17}$O($\alpha,\gamma$) rate (one tenth of the standard choice, i.e. \citealt{Caughlan88}). Since the only model produced by \citet{Fris16} with such assumptions is for a stellar mass of 25 $M_\odot$, we decided to compute for each element a ratio between the yields obtained from the fast rotator model and the previous one at 25 $M_\odot$, and then apply the resulting scale factors to the other models with metallicity [Fe/H] = $-3.8$ and masses 15, 20 and 40 $M_\odot$ (see \citealt{Cescutti13}).
\\The models developed by \citet{Fris16} and
their characteristics are reported in Table~\ref{tab:3}. Notice that, of the original models, we used only the ones with rotation.
\begin{table}
\centering
\footnotesize
\caption{Model parameters adopted for our work from \citet{Fris16}: initial mass, model label, initial ratio of surface velocity to critical velocity, time-averaged surface velocity during the MS phase, metallicity.}\label{tab:3}
\begin{tabular}{ccccc}
\hline
\hline
Mass (M{$_\odot$}) & Model & $v_{\text{ini}}/v_{\text{crit}}$ & $\langle v \rangle_{\text{MS}}$ (km/s) & [Fe/H] \\
\hline
15 & A15s4 & 0.4 & 200 & 0.0 \\
   & B15s4 & 0.4 & 234 & -1.8 \\
   & C15s4 & 0.4 & 277 & -3.8 \\
20 & A20s4 & 0.4 & 216 & 0.0 \\
   & B20s4 & 0.4 & 260 & -1.8 \\
   & C20s4 & 0.4 & 305 & -3.8 \\
25 & A25s4 & 0.4 & 214 & 0.0 \\
   & B25s4 & 0.4 & 285 & -1.8 \\
   & C25s4 & 0.4 & 333 & -3.8 \\
   & C25s5b$^a$ & 0.5 & 428 & -3.8 \\
40 & A40s4 & 0.4 & 186 & 0.0 \\
   & B40s4 & 0.4 & 334 & -1.8 \\
   & C40s4 & 0.4 & 409 & -3.8 \\
\hline
\multicolumn{5}{l}{$^a$ Models calculated with a lower $^{17}$O($\alpha,\gamma$).}
\end{tabular}
\end{table}
\\The work of \citet{Lim18} also produced new nucleosynthesis models
of massive stars, but with different assumptions. They considered a larger stellar mass range of 13-120 $M_\odot$, and four metallicities: [Fe/H] = 0, $-1$, $-2$ and $-3$. They produced a set of yields for stars with three possible velocities, using an initial speed of 0 km/s (non rotating), 150 km/s and 300 km/s. We decided here to develop three models assuming all stars rotate with the same initial speed. Realistically, a velocity distribution is expected for stellar rotation, but our assumption allows us to investigate a mean velocity.
\\Another important difference between the two papers is that in \citet{Lim18} the models have been computed up to the pre-SN stage, and the explosive nucleosynthesis has been taken into account by means of induced explosions, while \citet{Fris16} models stop at the beginning of the O-core burning. In the model we use, for \citet{Lim18} the amount of matter that effectively is ejected is the one lost by the star by stellar wind, during the pre-SN evolution, plus the one ejected during the explosion. The mass cut between the collapsing core and the ejected envelope has been fixed in such a way that the ejecta contains 0.07 $M_\odot$ of $^{56}$Ni, a typical value observed in the spectra of core-collapse SNe. In particular, from the \citet{Lim18} sets developed for this scenario, we used here the Set F, where each mass is considered to eject 0.07 $M_\odot$ of $^{56}$Ni.
\\A comparison between the models of the two works can be seen in Figure~\ref{fig:comp}.

\noindent Also for the r-process events, we investigate two
scenarios. In the first, MRD SNe are taken into account as r-process
events, as suggested in \citet{Winteler} and later confirmed by
\citet{Nishi}. In this case, we used the assumptions adopted in
\citet{Cescutti14}, where 10\% of all the stars in the range 10-80
$M_{\odot}$ are considered productive, with a timescale of 3-25 Myr. In
the second, we study the possibility of using NSMs instead of MRD SNe
as r-production sites \citep{Ross}. In this case, the rate and the
yields were adopted from the work of \citet{Matteucci} and
\citet{Cescutti15} respectively. They suggest that r-element material
can be produced only by NSMs if the neutron stars originate in the
stellar mass range of 9-50 $M_\odot$, the coalescence timescale is
fixed and equal to 1 Myr and the rate of NSM events is 0.018, the one
of \citet{Kalo04}. The recent observations of the LIGO/Virgo rate for
the event GW170817 seem to confirm this result \citep{Matteucci18}.
In both r-process cases, the adopted scaling factor between Sr and Ba
has been taken from the solar system r-process contribution as
determined by \citet{SSC04}.  This simple approach is reasonable,
since the emphasis of this paper is not on the r-process issue but on
the contribution from rotating massive stars to the production of
strontium and barium.
\\Finally, concerning the iron yields from core-collapse SNe, we decided to adopt the ones from \citet{Kobayashi06}, the same used by \citet{Matteucci}, instead of the ones from 
\citet{Lim18}. The reason for this choice is that the two works approach very similar results, as we chose to be consistent with \citet{Matteucci}.

\section{Results}
\subsection{Ratios of heavy elements}
\label{subsec:5.1}
We present here the ratios for [Ba/Fe], [Sr/Fe] and [Sr/Ba]. The
reason for this choice is to study the trends of production for
strontium and barium as representative respectively of the first and
second peak of the s-process production. Moreover, the behaviour of
[Sr/Ba] can provide a differential information about the production of
these elements by the s-process in rotating massive stars, that is the
focus of the present work.  \\In
order to make the trend of the data clearer, the metallicity range has
been divided into equal sections; for each one, a mean value of
abundance for observational data has been computed. For every mean dot
an error has been associated, estimated as standard deviation; the set
of error bars delimits a grey shadowed area which we consider the
acceptable zone.  \\We run the models using the prescriptions
described in Section~\ref{sec:sec4}. The models we developed and their
features are summarized in Table~\ref{tab:4}.
\begin{table*}
\centering
\footnotesize
\caption{The models and their prescriptions.}\label{tab:4}
\begin{tabular}{cccc}
\hline
\hline
Model name&s-process in rotating massive stars&Assumed stellar rotational velocity&r-process\\
\hline
F+MRD & \citet{Fris16} & see Table~\ref{tab:3} & \citet{Ross, Winteler} \\
LC000+MRD & \citet{Lim18} & non rotating & \citet{Ross, Winteler} \\
LC150+MRD & \citet{Lim18} & 150 km/s & \citet{Ross, Winteler} \\
LC300+MRD & \citet{Lim18} & 300 km/s & \citet{Ross, Winteler} \\
LC075+MRD & \citet{Lim18} & 75 km/s$^a$ & \citet{Ross, Winteler} \\
LC225+MRD & \citet{Lim18} & 225 km/s$^a$ & \citet{Ross, Winteler} \\
F+NSM & \citet{Fris16} & see Table~\ref{tab:3} & \citet{Matteucci, Cescutti15} \\
LC000+NSM & \citet{Lim18} & non rotating & \citet{Matteucci, Cescutti15} \\
LC150+NSM & \citet{Lim18} & 150 km/s & \citet{Matteucci, Cescutti15} \\
LC300+NSM & \citet{Lim18} & 300 km/s & \citet{Matteucci, Cescutti15} \\
\hline
\multicolumn{4}{l}{$^a$ Yields obtained by interpolation process: see Section~\ref{subsec:5.2}}
\end{tabular}
\end{table*}
\begin{figure*}
\centering
\footnotesize
\includegraphics[scale=0.3]{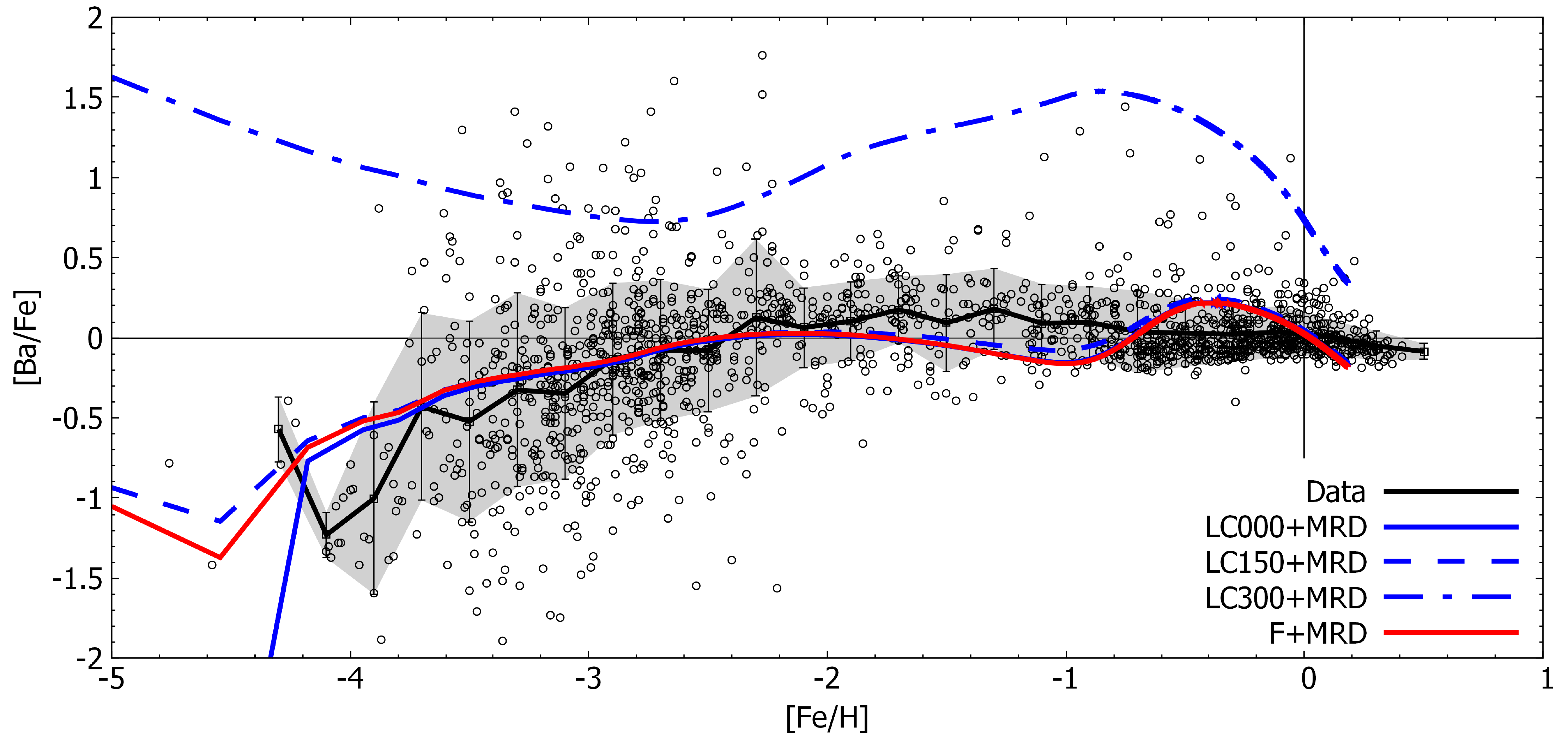}
 \caption{[Ba/Fe] versus [Fe/H]. The black dots, track and shadowed area are the observations (sources listed in Table~\ref{tab:1}); red line is model F+MRD; blue solid line is model LC000+MRD; blue dashed line is model LC150+MRD; blue double dashed line is model LC300+MRD (see Table~\ref{tab:4}).}
 \label{fig:2}
\end{figure*}
\begin{figure*}
\centering
\footnotesize
 \includegraphics[scale=0.3]{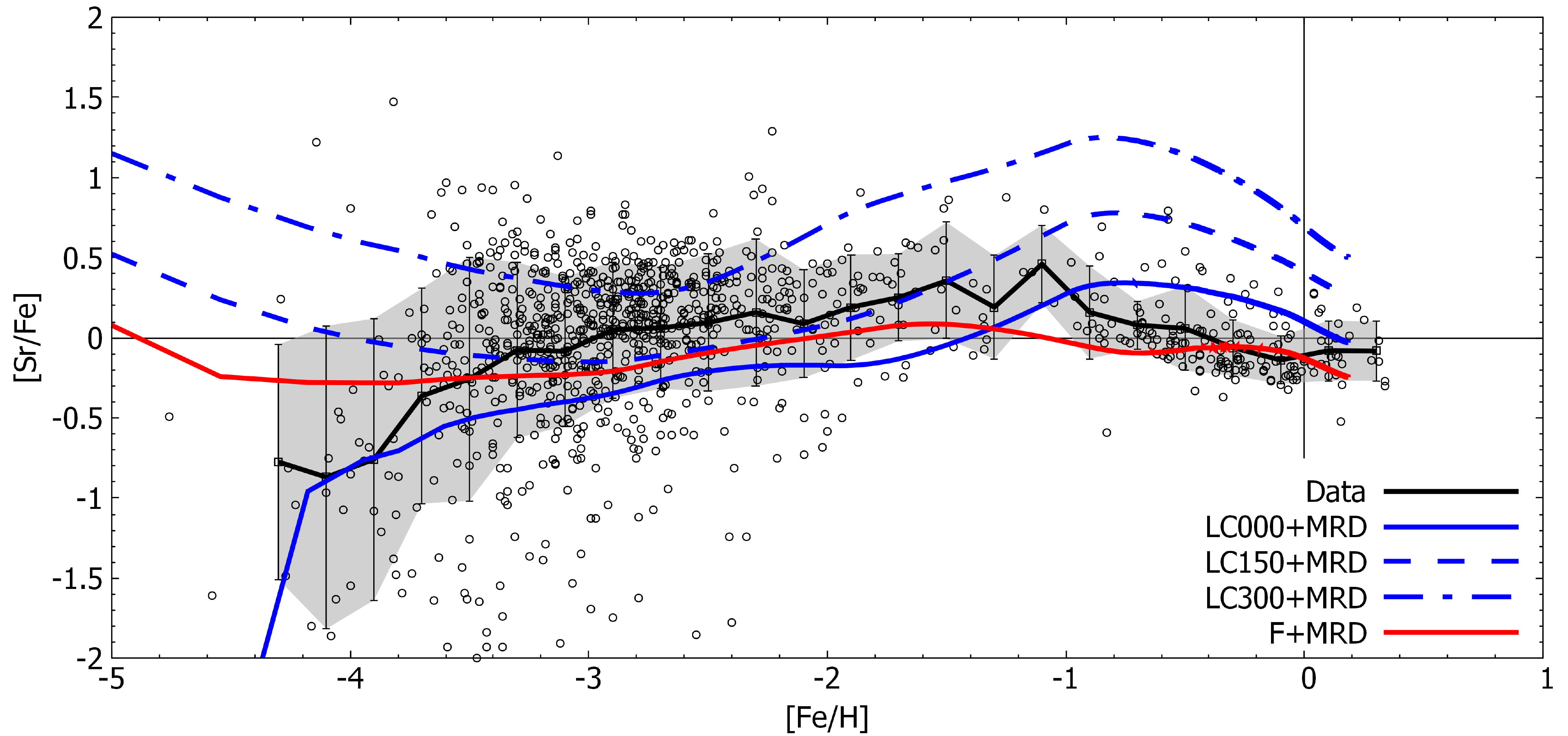}
 \caption{[Sr/Fe] versus [Fe/H]. The black dots, track and shadowed area are the observations (sources listed in Table~\ref{tab:1}); red line is model F+MRD; blue solid line is model LC000+MRD; blue dashed line is model LC150+MRD; blue double dashed line is model LC300+MRD (see Table~\ref{tab:4}).}
 \label{fig:3}
\end{figure*}
\begin{figure*}
\centering
\footnotesize
 \includegraphics[scale=0.3]{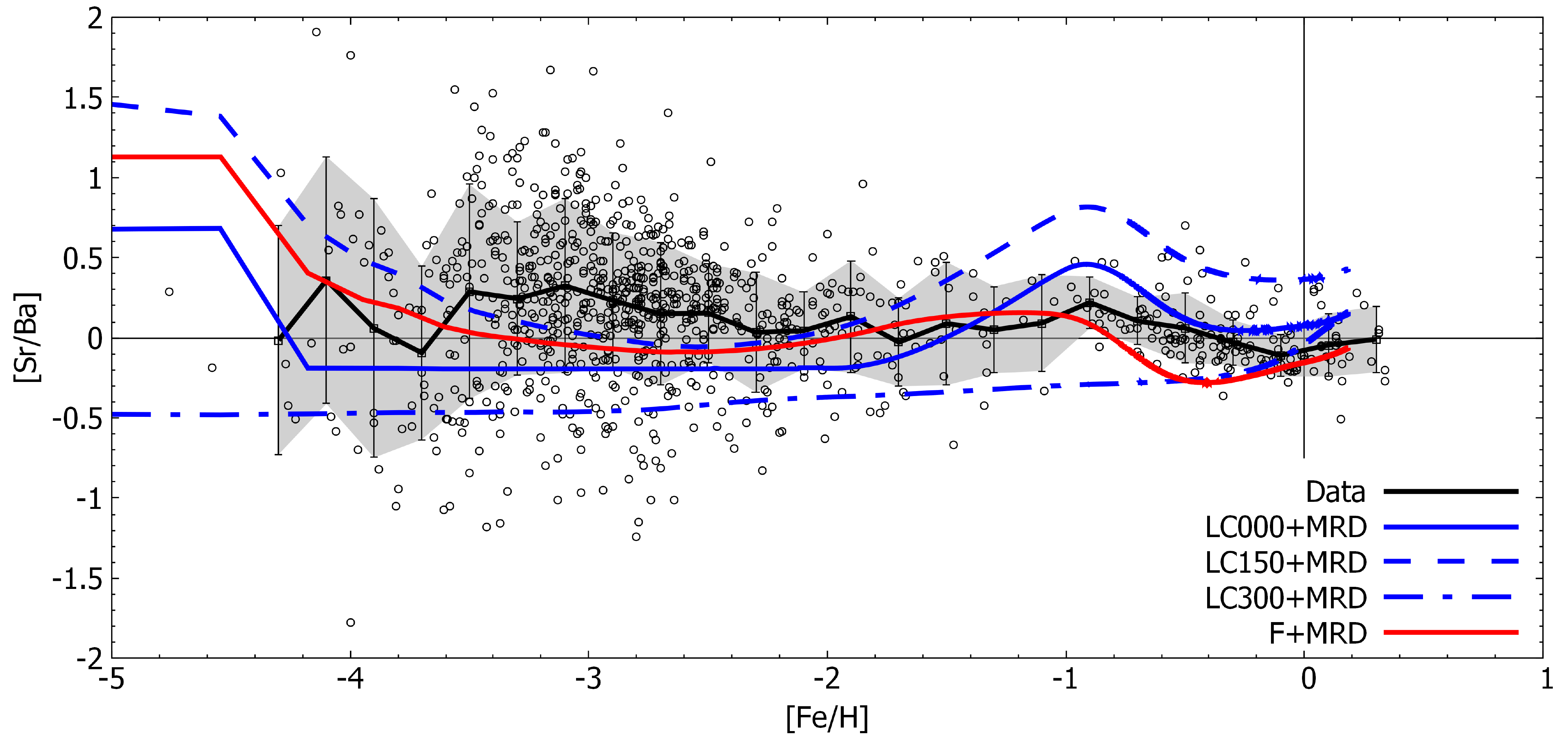}
 \caption{[Sr/Ba] versus [Fe/H]. The black dots, track and shadowed area are the observations (sources listed in Table~\ref{tab:1}); red line is model F+MRD; blue solid line is model LC000+MRD; blue dashed line is model LC150+MRD; blue double dashed line is model LC300+MRD (see Table~\ref{tab:4}).}
 \label{fig:4}
\end{figure*}
\begin{figure*}
\centering
\footnotesize
 \includegraphics[scale=0.3]{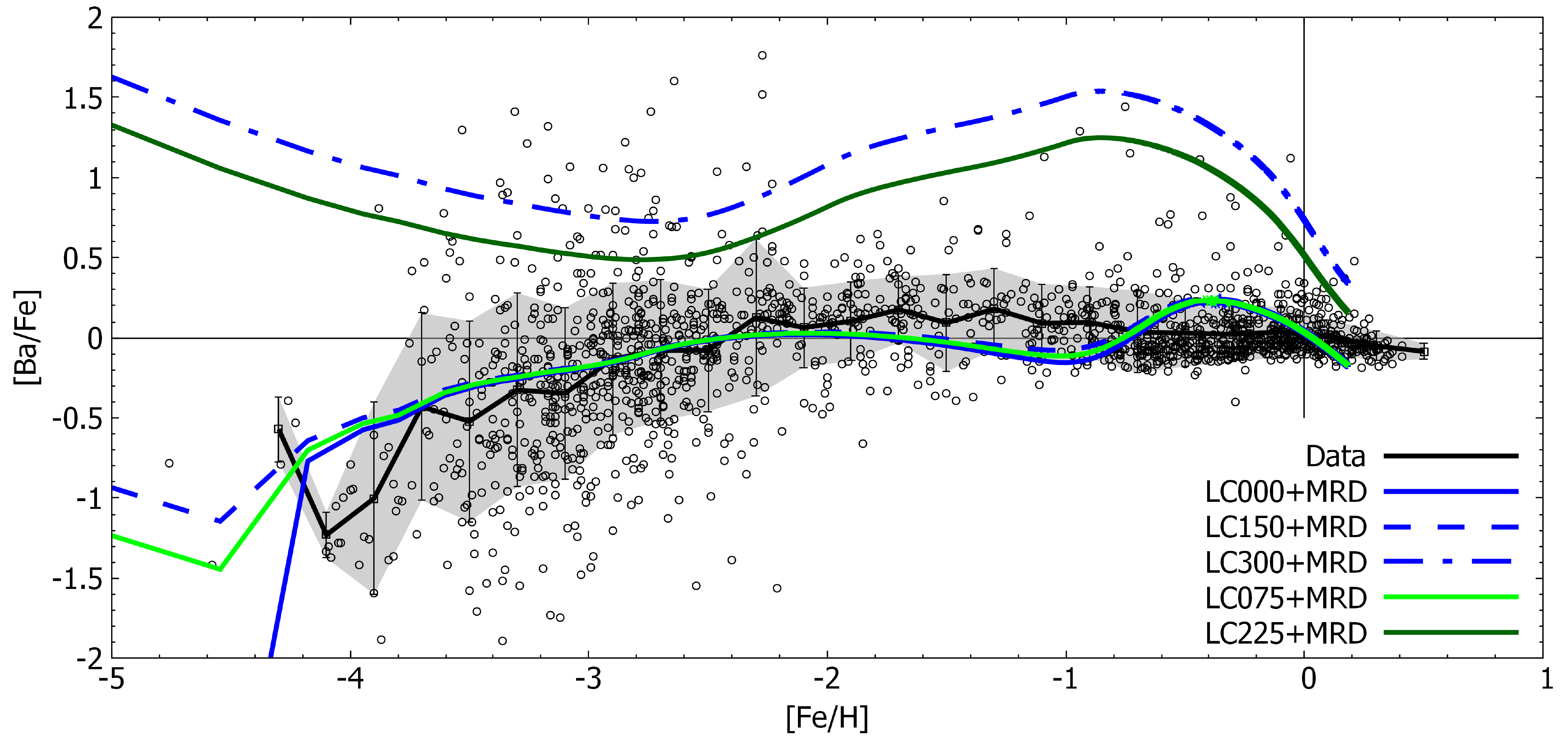}
  \caption{[Ba/Fe] versus [Fe/H]. The black dots, track and shadowed area are the observations (sources listed in Table~\ref{tab:1}); blue solid line is model LC000+MRD; blue dashed line is model LC150+MRD; blue double dashed line is model LC300+MRD; light green line is model LC075+MRD; dark green line is model LC225+MRD (see Table~\ref{tab:4}).}
 \label{fig:5}
\end{figure*}
\\First we present the results obtained with MRD SNe as r-process
sites. The ratio of [Ba/Fe] versus [Fe/H] is shown in
Figure~\ref{fig:2}. The dots represent the observational data: for
them, a black mean trend is tracked. The red line is model F+MRD with
\citet{Fris16} yields, the blue lines are relative to models LC000+MRD
(solid), LC150+MRD (dashed) and LC300+MRD (double dashed), using \citet{Lim18} yields assuming all stars rotate with the same initial velocity 0 km/s, 150 km/s and 300 km/s  respectively. \\As for the two prescriptions, we can see that the results obtained with the use of the F, LC000 and LC150 set of yields cover properly the data in every range of metallicity; the use of the LC300 yields, on the contrary, produce an amount of barium that is not compatible with the observations. The reason for this result is that barium is mostly
produced by r-process in our framework, so in the case of extremely
high production by s-process, our model is not compatible with the
data.\\ We recall that the s-process from AGB stars has a long
timescale, so in the earliest stages its production is not
significant. We still have the s-process from rotating massive stars,
but such a contribution is in general lower than the one from the
r-process.  \\Around [Fe/H] $=-1$, the contribution for barium by AGB
stars starts to be effective; its effect balances the production of
iron by SNe Ia: the resulting trend is almost flat.  \\We can observe
a decline also toward metallicities higher than solar. The recent work
by \citet{Prantzos} predict this behaviour for barium and the second
peak s-elements too; the chemical abundances of stars with super solar
metallicity show a decline too, but less extreme than model
predictions. \\Similar results can be found for the [Sr/Fe] ratio,
reported in Figure~\ref{fig:3}. F, LC000 and LC150 yields match remarkably well the data. On the other hand, we found that the results of the model with all stars rotating at the highest velocity (LC300) overproduce the ratio of [Sr/Fe]. A light decline toward metallicities higher than solar is seen
but, unlike the previous case, observations do not show such a trend
for strontium. For strontium all the models with rotations by
\citet{Lim18} overestimate the solar abundance, contrary to barium
models.  \\Finally, the key ratio - [Sr/Ba] - is presented in
Figure~\ref{fig:4}. For this ratio, the chemical abundances of stars at
$-4<\text{[Fe/H]}<-3$ are on average (black line) higher than the
[Sr/Ba] ratio by r-process only.  For the r-process contribution only,
we can compare to model LC000+MRD at these metallicities, which is
almost exclusively r-process; in fact with no rotation, there is no
s-process production of Sr and Ba. We also note that for [Fe/H]$<-$4,
it is difficult to claim any clear trend, given also the scarse number
of data. 
\\For this plot, assuming the low velocity yields (150 km/s) from \citet{Lim18} for all stars returns the most accurate trend which satisfies the observations. As we said in Sec~\ref{sec:sec4}, this is an extreme assumption, for it does not allow us to appreciate the full distribution of velocities. The non-rotating yields or the high velocity ones represent extremities of such distribution; there are certainly some stars which possess these features, but the data tell they are not frequent.
The reasons for this behaviour are different: in the first case of no rotation, with no
production by s-process, the r-process enrichment do not fit the data;
in the second of high speed rotation, the model produces too much Ba compared
to Sr. \\On the other hand, the low velocity set displays the best
behaviour compared to the data, although possibly the high value of
[Sr/Ba] in the model is present at [Fe/H] $\sim -$4, with a 0.5 dex
displacement compared to the data at [Fe/H] $\sim -$3.5. This can be
due to the single chemical evolution model that we use to interpret the
data from the Galactic halo to the solar metallicity.  At this low
metallicity the model obtained using the nucleosynthesis computed in
\citet{Fris16} performed similarly to the model assuming 150 km/s from
\citet{Lim18}. We interpret this due to the s-process contribution
of rotating massive stars that was also the main conclusion in
\citet{Cescutti13}, that was based only on the yields by
\citet{Fris12}.

\begin{figure*}
\centering
\footnotesize
 \includegraphics[scale=0.3]{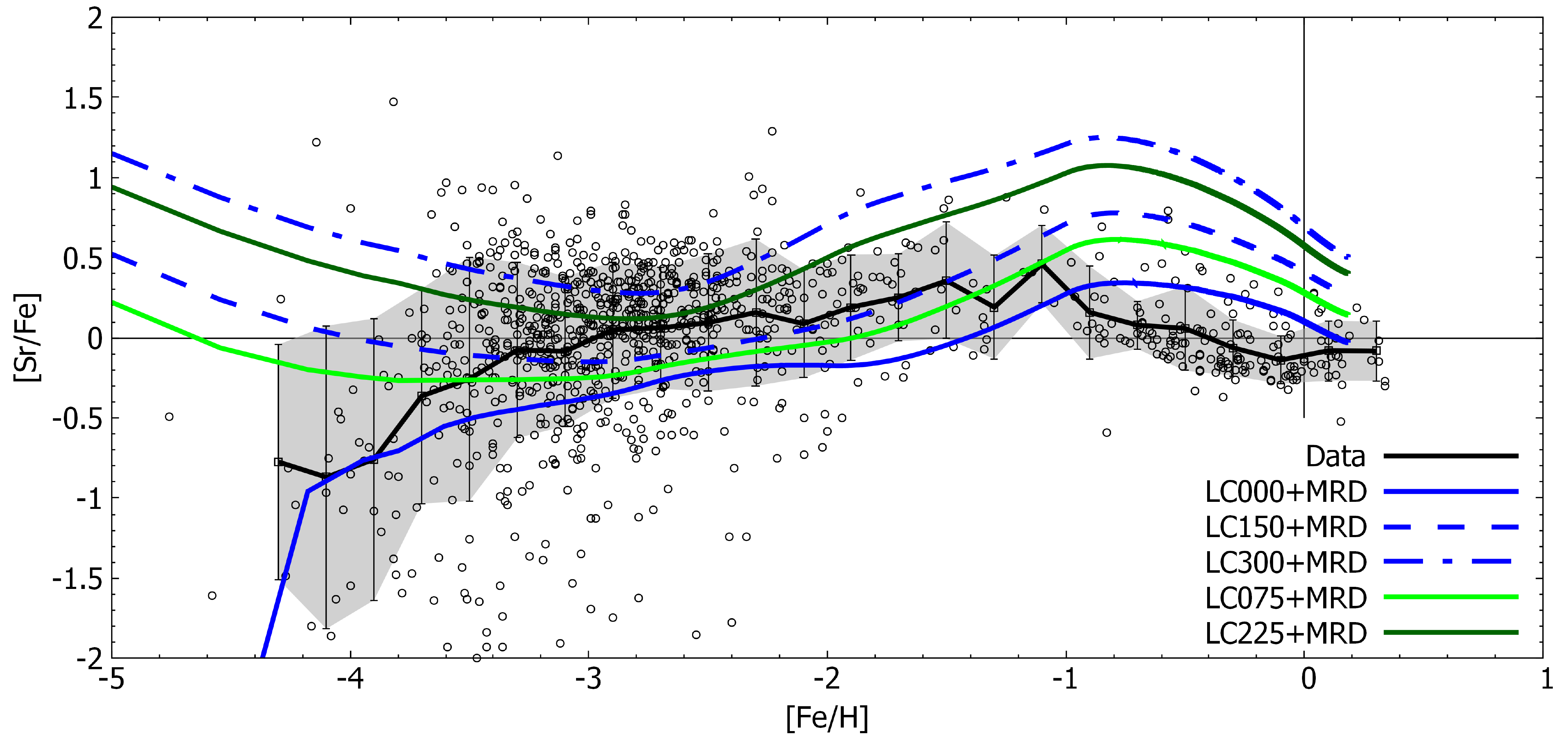}
  \caption{[Sr/Fe] versus [Fe/H]. The black dots, track and shadowed area are the observations (sources listed in Table~\ref{tab:1}); blue solid line is model LC000+MRD; blue dashed line is model LC150+MRD; blue double dashed line is model LC300+MRD; light green line is model LC075+MRD; dark green line is model LC225+MRD (see Table~\ref{tab:4}).}
 \label{fig:6}
\end{figure*}
\newpage
\begin{figure*}
\centering
\footnotesize
 \includegraphics[scale=0.3]{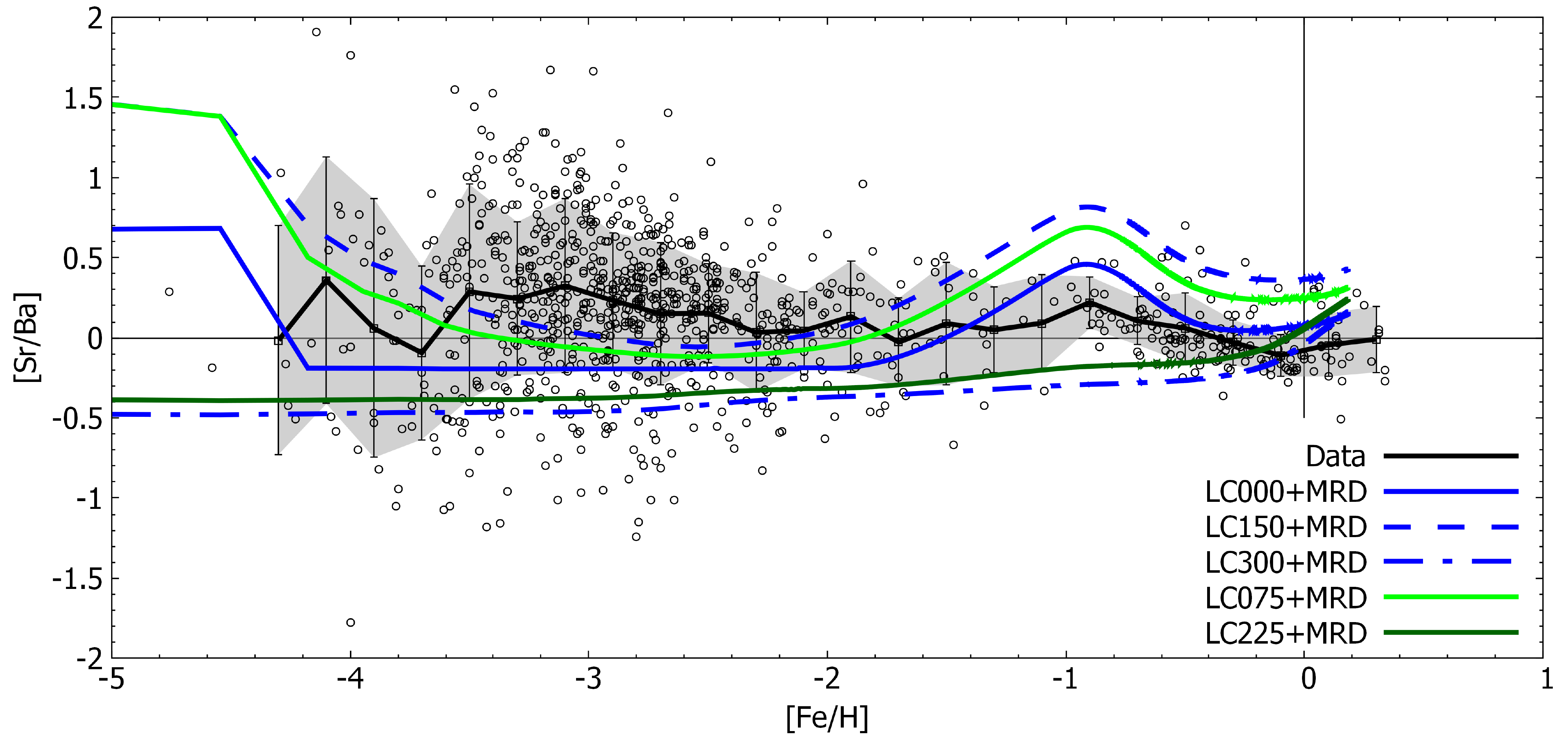}
  \caption{[Sr/Ba] versus [Fe/H]. The black dots, track and shadowed area are the observations (sources listed in Table~\ref{tab:1}); blue solid line is model LC000+MRD; blue dashed line is model LC150+MRD; blue double dashed line is model LC300+MRD; light green line is model LC075+MRD; dark green line is model LC225+MRD (see Table~\ref{tab:4}).}
 \label{fig:7}
\end{figure*}

\subsection{Velocity interpolation}
\label{subsec:5.2}
Considering the results of the previous section \ref{subsec:5.1}, we decided to
investigate rotational velocities intermediate between those in the
grid proposed by \citet{Lim18}.  We used a linear interpolation over
mass and metallicity on two prescriptions different in stellar speed, to obtain a
new set with intermediate features, which we denoted by the arithmetic
mean velocity. Two interpolations had been produced: 75 km/s between
the sets 0 and 150 km/s (model LC075+MRD), and 225 km/s between 150
and 300 km/s (model LC225+MRD).  \\The results of such method for
[Ba/Fe], [Sr/Fe] and [Sr/Ba] are displayed in
Figure~\ref{fig:5},~\ref{fig:6}, and~\ref{fig:7} respectively, with a
light green line for model LC075+MRD and a dark green one for model
LC225+MRD, compared with the original blue models LC000+MRD, LC150+MRD
and LC300+MRD. The new trends are, as we expected, intermediate between the original tracks. 
\\A comparison between models is useful to determine which velocity better covers different scenarios: we can identify the best-fitting model for a range of metallicity, therefore the rotational velocity of most stars in that range.
\\In particular, from the [Sr/Fe] and [Sr/Ba] behaviour, we can see that up to [Fe/H] $=-2$, the mean velocity which best covers the data is 150 km/s; from [Fe/H] $=-2$ to $-1$, a transition to 75 km/s can be seen, while for [Fe/H] $>-1$ the non rotating assumption is preferred. \\In \citet{Prantzos}, where we recall the same \citet{Lim18} data were used, a similar assumption was made: the initial velocity of massive stars is on average about 180 km/s up to [Fe/H] $=-3$, then it decreases until reaching a plateau of 50 km/s at solar metallicities. As we can see, our predicted velocities are in general lower.
\begin{figure*}
\centering
\footnotesize
 \includegraphics[scale=0.3]{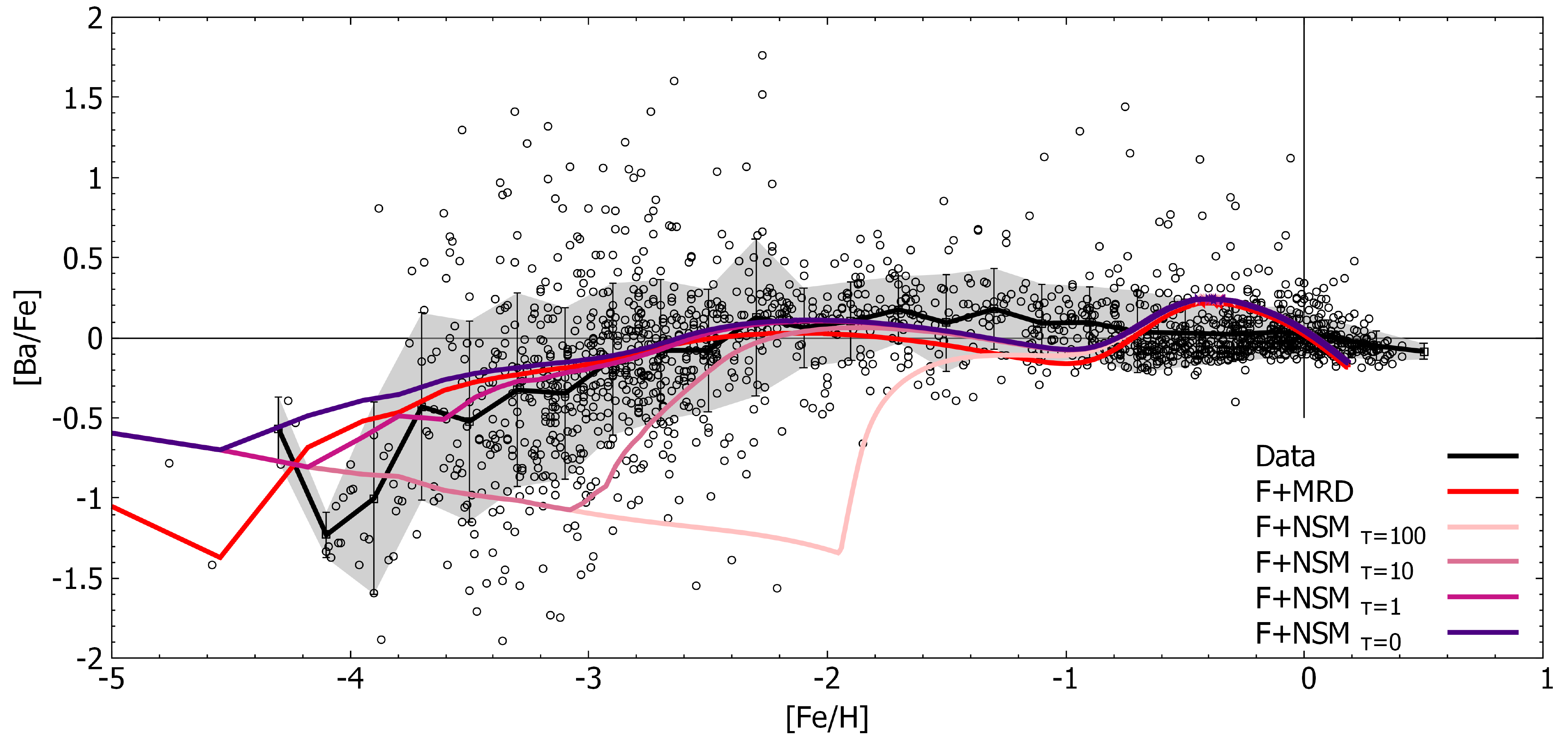}
   \caption{[Ba/Fe] versus [Fe/H]. The black dots, track and shadowed area are the observations (sources listed in Table~\ref{tab:1}); red line is model F+MRD; purple lines are model F+NSM with variations in the time delay, namely (from darker to lighter) $\tau=0,1,10,100$ Myr (see Table~\ref{tab:4}).}
 \label{fig:8}
\end{figure*}

\subsubsection{A mixed model}
In Section~\ref{subsec:5.1}, from the study on [Sr/Ba] ratio at low metallicities, the scenario provided by \citet{Lim18} indicates that non-rotating stars cannot explain the chemical abundances in the Galactic halo around [Fe/H] $=-3$, but the best fit of the data assumes stars having on average the low velocity 150
km/s or the interpolation 75 km/s.  On the other
hand, the non-rotating case fits the data at solar metallicities quite
well, while the model with rotation at 150 km/s (as well as the
model at 75 km/s) predicts a trend with a too high [Sr/Ba] ratio
compared to the data.  We can conclude that a good fit could be
produced by a model using the rotating \citet{Lim18} prescriptions
for low metallicities up to [Fe/H] $\sim -$2, and the non-rotating ones at higher metallicities.  \\Such a hypothesis is in agreement
with the assumptions taken by the Geneve group
\citep{Meynet06,Hirschi07}
and adopted also by \citet{Fris16}. They assume in fact a constant
ratio between initial velocity and critical velocity. This produces
automatically
an higher velocity for models at low metallicity, as we
reported in Section~\ref{sec:sec4} (see Table~\ref{tab:3}).  
\\As we noticed, the same conclusion has been suggested within the paper of \citet{Prantzos}, but they based their constraint on nitrogen trend.
\\Assuming no rotation (or low rotational velocity, below 75 km/s) is a specific result of our models LC. For example, in the results considering yields from \citet{Fris16} (see Table~\ref{tab:3}), a minimum rotation is still assumed (and required) to explain observations at best. 
\\Moreover, this outcome is driven only by considering neutron capture elements; indeed, for nitrogen, \citet{Prantzos} obtained slightly different constraints.
\begin{figure*}
\centering
\footnotesize
 \includegraphics[scale=0.3]{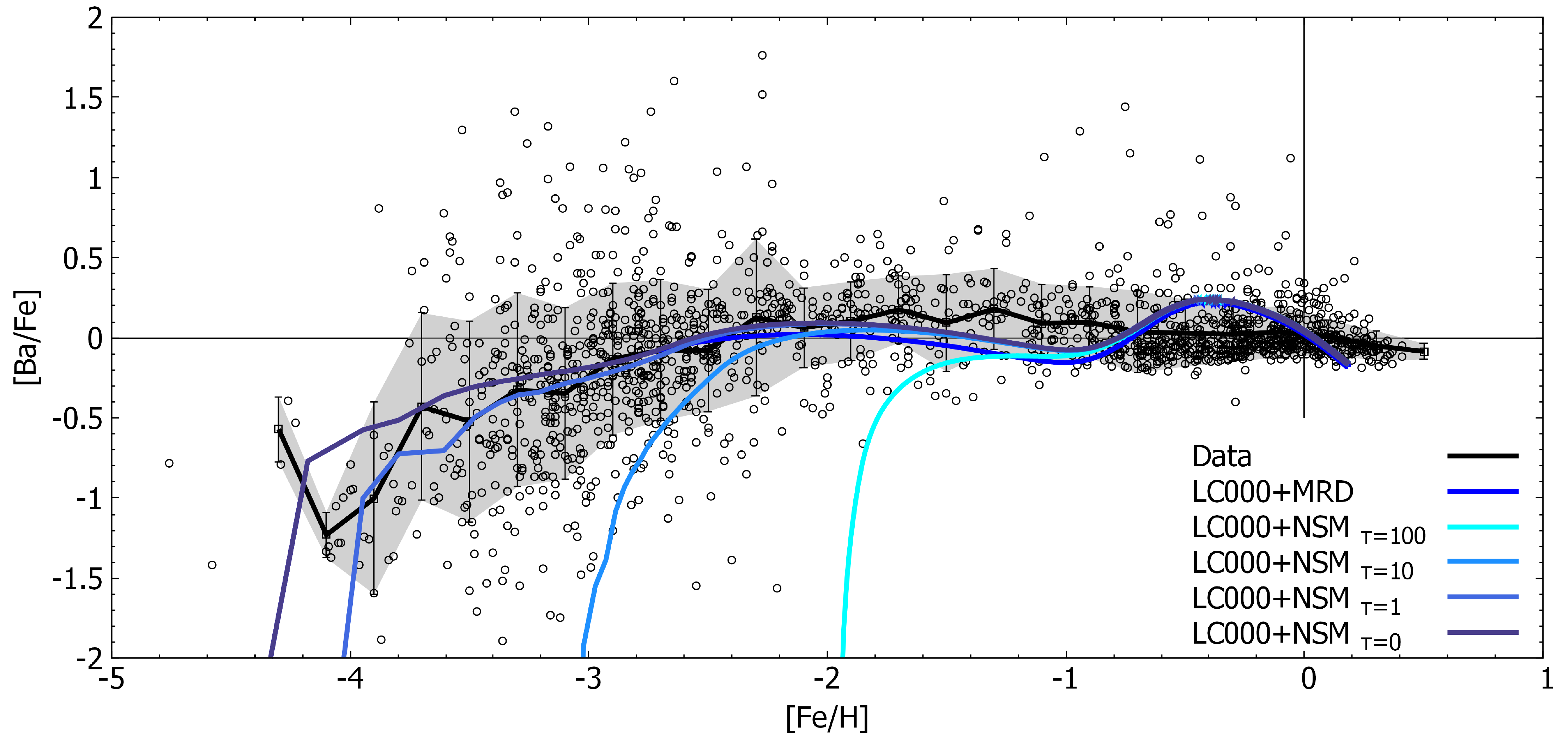}
    \caption{[Ba/Fe] versus [Fe/H]. The black dots, track and shadowed area are the observations (sources listed in Table~\ref{tab:1}); dark blue line is model LC000+MRD; lighter blue lines are model LC000+NSM with variations in the time delay, namely (from darker to lighter) $\tau=0,1,10,100$ Myr (see Table~\ref{tab:4}).}
 \label{fig:9}
\end{figure*}
\begin{figure*}
\centering
\footnotesize
 \includegraphics[scale=0.3]{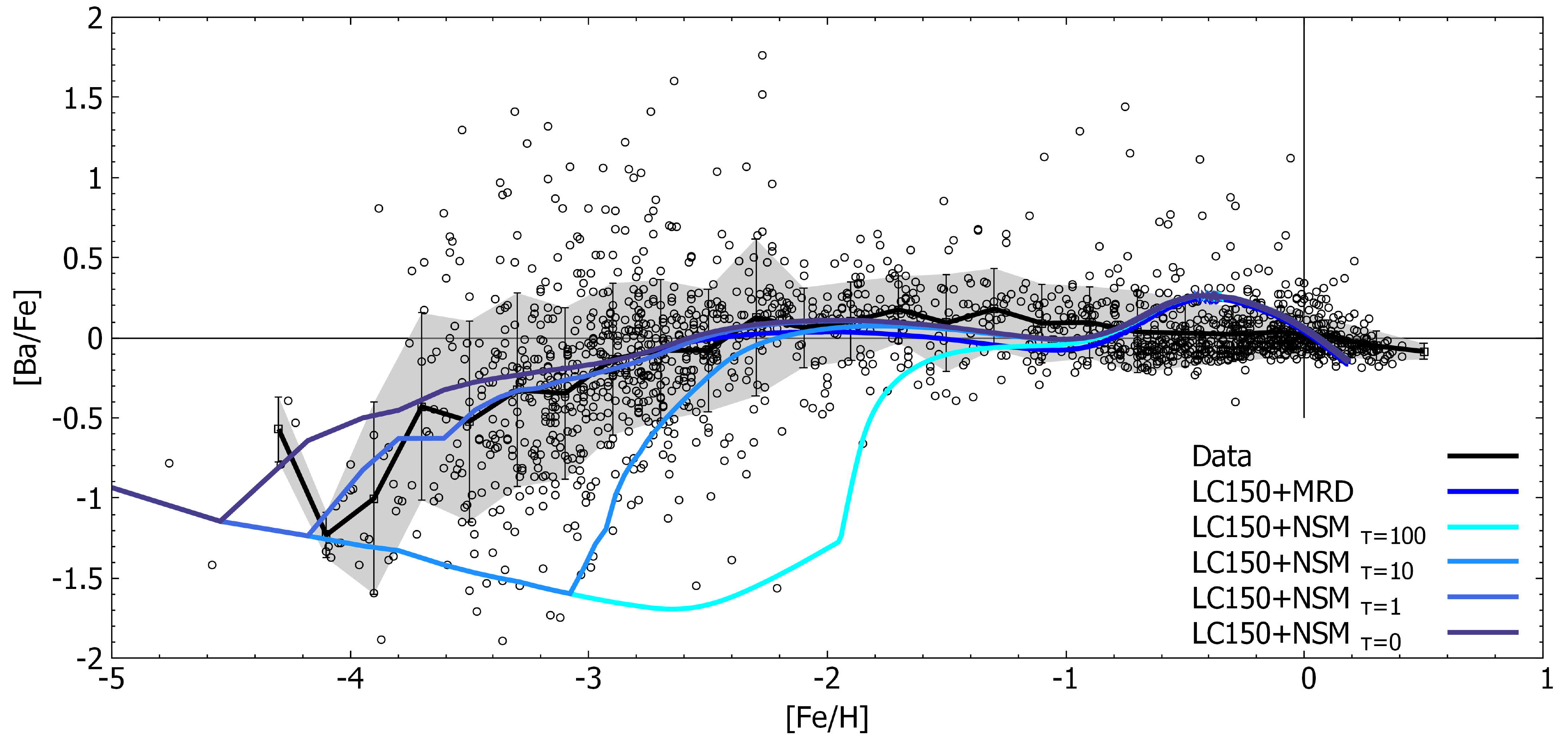}
    \caption{[Ba/Fe] versus [Fe/H]. The black dots, track and shadowed area are the observations (sources listed in Table~\ref{tab:1}); dark blue line is model LC150+MRD; lighter blue lines are model LC150+NSM with variations in the time delay, namely (from darker to lighter) $\tau=0,1,10,100$ Myr (see Table~\ref{tab:4}).}
 \label{fig:10}
\end{figure*}
\begin{figure*}
\centering
\footnotesize
 \includegraphics[scale=0.3]{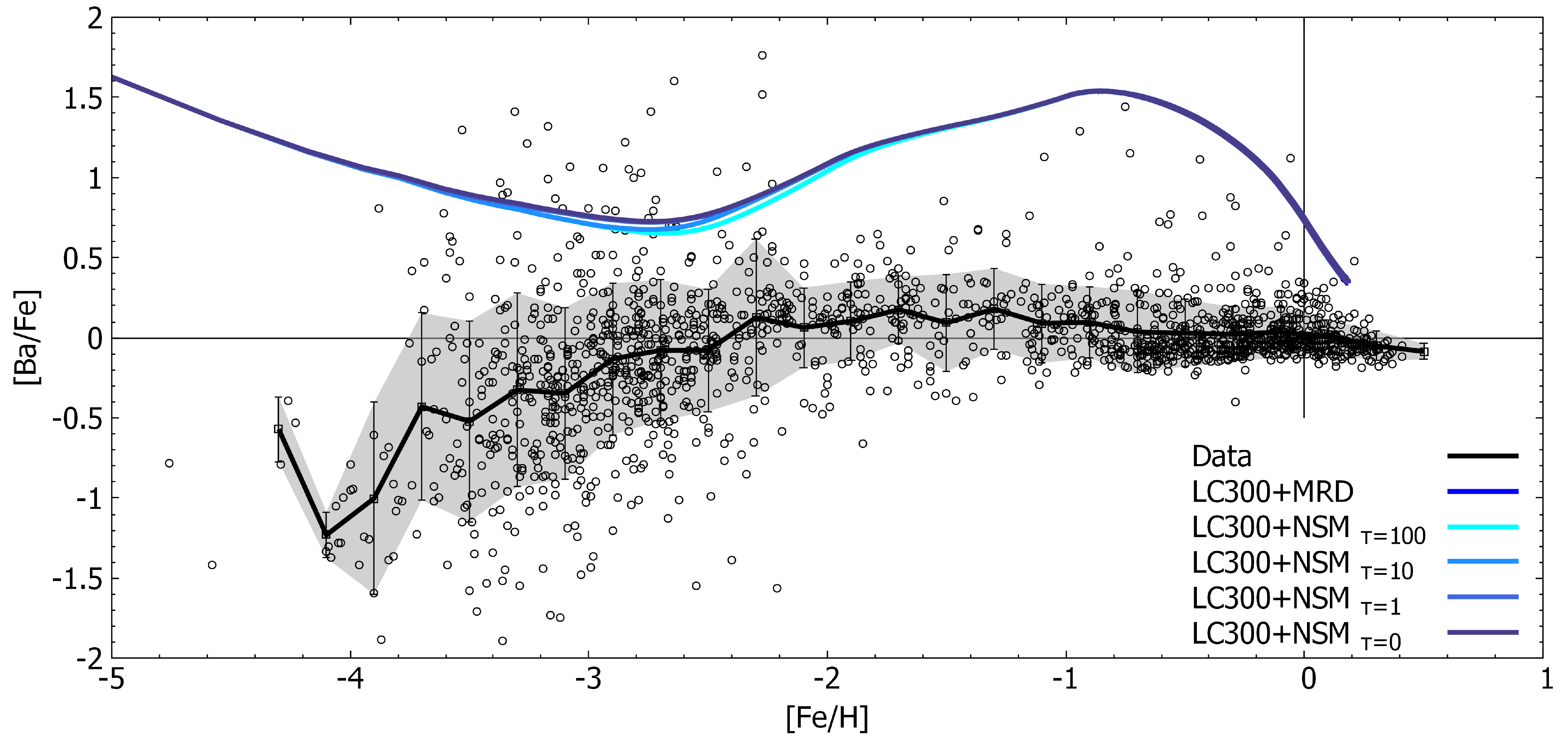}
    \caption{[Ba/Fe] versus [Fe/H]. The black dots, track and shadowed area are the observations (sources listed in Table~\ref{tab:1}); dark blue line is model LC300+MRD; lighter blue lines are model LC300+NSM with variations in the time delay, namely (from darker to lighter) $\tau=0,1,10,100$ Myr (see Table~\ref{tab:4}).}
 \label{fig:11}
\end{figure*}

\subsection{Neutron star mergers and time delay}
We decided to verify the impact of using NSMs as r-process sites
instead of MRD SNe. For this test, we need also to assume a
coalescence timescale for the NSMs. We investigate these timescales:
$\tau = 0,\ 1,\ 10$ or 100 Myr. In this model we assume that the Ba
and Sr fractions which originate from the r-process are formed in
NSMs; in particular, the yield of Ba and Sr are obtained by scaling
the Eu contribution according to the abundance ratios observed in
solar abundance (see also Sect. \ref{sec:sec4}) \\In Fig.~\ref{fig:8},
we present the models F+NSM using \citet{Fris16} yields for massive
star s-process and NSMs for r-process: the trends, for different NSM
time-delays, are in purple (with longer $\tau$ presented with lighter
shades); in red, model F+MRD with \citet{Fris16} yields and MRD
SNe. From Fig. ~\ref{fig:8}, it is evident that the best choice for
the coalescence timescale is the shortest timescale: $\tau = 1$ Myr.
It is also noticeable that using \citet{Fris16} yields and NSMs for
$\tau = 0$ or 1 Myr is very similar to the previous model with MRD
SNe.  \\An extensive analysis of the best choice for the delay time
can be found in \citet{Matteucci} and \citet{Cescutti15}; also these works
conclude that NSMs with a coalescence timescale of 1 Myr can be a
reasonable scenario concerning the europium production in the Galaxy.
\\Also \citet{Lim18} yields has been tested together with NSMs:
Fig.s ~\ref{fig:9} and~\ref{fig:10} display the modified models
LC000+NSM and LC150+NSM with the non rotating and the 150 km/s data
sets respectively ($\tau$ increases with lighter blue shades); in dark
blue, models with MRD SNe. The conclusions are similar to the previous
case: the best choice for NSMs is a time delay of 1 Myr, while with a
coalescence time of $\tau=0$ Myr the trend results very similar to the
model where MRD SNe are used.  \\Finally, Figure~\ref{fig:11} shows
the models LC300+NSM with rotating 300 km/s \citet{Lim18} yields. In
this case, the higher velocity adopted produces 
trends that are not compatible with the abundances measured in
Galactic halo stars. Concerning fast rotation, a variation of NSM
coalescence timescale does not seem to affect the behaviour in any
way: the trends for fastest speeds appear unaltered.

\section{Conclusions}

In this paper, the nucleosynthesis of neutron capture elements has
been studied; we focused our attention on the production of the heavy
elements barium and strontium. Stellar rotation has been taken into
account: recent studies showed that it has a deep impact on stellar
nucleosynthesis, enhancing heavy element production.  \\In this work,
prescriptions from different authors have been tested and compared
with the most recent observational data, in order to verify the
differences and the similiarities by means of a chemical evolution
model.  The results were compared with the observational data to
verify which assumptions reproduced better the observations. We
compared the prescriptions assumed for s-process component in rotating
massive stars taken from \citet{Fris16} and from  \citet{Lim18}, in which three different initial rotational speeds where considered. We
also tested two possible sources of r-process material, namely NSMs,
assuming the yields from \citet{Matteucci} and MRD SNe, using the
prescriptions from \citet{Cescutti14}.  \\We summarize our conclusions
as follows.
\begin{itemize}
\item The best fits to tha data were the ones  with the yield set of \citet{Fris16} or assuming all stars rotate with an initial velocity of 150 km/s from \citet{Lim18}.
\item Although the non rotating models could explain the average trends of [Sr/Fe] and [Ba/Fe], so the pure production of
  r-process for these elements,
  they produce results non compatible for the [Sr/Ba] ratio, proving
  that rotation is a necessary assumption to reproduce the actual
  behaviour of the observational data.
\item Interpolations over the \citet{Lim18} velocity sets have been
  calculated. The model assuming a velocity of 75 km/s has shown to produce results compatible with the data. 
  \item Using the nucleosynthesis by \citet{Lim18} the best fit to the
  data could be obtained by assuming the slow rotating yields at the lowest
  metallicities and the non rotating yields for higher metallicities.
\item The models with NSMs as r-process sources have been tested with
  a NSM time delay of 0, 1, 10 and 100 Myr. Both \citet{Fris16} and
  \citet{Lim18} prescriptions for the low velocity set
  displayed an optimal behaviour with $\tau=1$ Myr. The variation of
  the r-process site does not change our conclusions concerning the
  rotating massive stars. 
\end{itemize}
We conclude that \citet{Fris16} assumption of different rotational
velocity as function of metallicity is valid, since it is effective in
reproducing the observations in every scenario.  \\On the other hand,
we can use the nucleosynthesis by \citet{Lim18}, but
different velocities should be adopted to fit at best the data;
in particular assuming an initial velocity of 150 km/s at low
metallicity and no rotation or very low rotation above
[Fe/H] $\sim -$2. A a similar conclusion has been reached in
\citet{Prantzos}, studying the nitrogen production.
\\However, the comparison to \citet{Prantzos} is to be carefully considered: their work used \citet{Lim18} yields for massive stars, but different assumptions were made. In particular, stars larger than 25 $M_\odot$ fail to explode and always fall back into black holes, contributing to the chemical
enrichment only through stellar winds. Furthermore, a direct observation of barium behaviour at low metallicities shows that \citet{Prantzos} employed a stronger s-process from rotating massive stars, for they assumed the average velocity to depend on metallicity, faster at low [Fe/H], while in our framework Ba is mostly produced by r-process.
\\Our study of
neutron star mergers have shown they are a good source for r-process:
the best fit is obtained for a coalescence time of 1 Myr. The same
conclusions have been achieved in the works of \citet{Matteucci} and
\citet{Cescutti15}.

\section*{Acknowledgements}
GC acknowledges financial support from the EU Horizon2020 programme under the Marie
Sklodowska-Curie grant 664931. FM acknowledges funds from University of Trieste (Fondo per la Ricerca d'Ateneo - FRA2016). 
RH acknowledges support from the World Premier International Research Center Initiative (WPI Initiative), MEXT, Japan. GC and RH acknowledge support from the ChETEC COST Action (CA16117), supported by COST (European Cooperation in Science and Technology).
This work has been partially supported by the Italian grants ``Premiale 2015 MITiC'' (P.I. B. Garilli) and ``Premiale 2015 FIGARO'' (P.I. G. Gemme).




\bibliographystyle{mnras}
\bibliography{article} 


\bsp	
\label{lastpage}
\end{document}